\documentstyle[12pt]{article}

\begin{document}
\renewcommand{\thefootnote}{\fnsymbol{footnote}}
\newpage
\setcounter{page}{0}

\phantom{
{January 1997}\\
{JINR E2-97-21}\\
{solv-int/9701019}
}

\vfill

\begin{center}
{\LARGE {\bf To the Gel'fand-Tsetlin realization of }}\\ [0.3cm]
{\LARGE {\bf irreducible representations of classical }}\\ [0.3cm]
{\LARGE {\bf semisimple algebras}}\\ [1cm]
{\large  A.N. Leznov$^{a,b}$ }\\
{\em  Instituto de Investigaciones en Matem\'aticas
  Aplicadas y en Sistemas}~\\
{\em Universidad Nacional Aut\`onoma de M\`exico} \\
{\em Apartado Postal 48-3, 62251 Cuernavaca,    \\
     Morelos, M\'exico} \\
\quad \\
{\em {~$~^{(a)}$ On leave from Institute for High Energy Physics,}}\\
{\em 142284 Protvino, Moscow Region, Russia}\\
{\em {~$~^{(b)}$ Also at Bogoliubov Laboratory of Theoretical Physics, 
JINR,}}\\
{\em 141980 Dubna, Moscow Region, Russia}~\quad\\

\end{center}

\vfill

\centerline{ {\bf Abstract}}

It is shown that the Gel'fand-Tsetlin realization of irreducible 
representations of the $A_n$ algebra is directly connected with 
a linear exactly integrable system in the n-dimensional space. 
General solution for this system is explicitly given.

\vfill
{\em e-mail:\\
1) leznov@ce.ifisicam.unam.mx }

\newpage
\pagestyle{plain}
\renewcommand{\thefootnote}{\arabic{footnote}}
\setcounter{footnote}{0}

\section{Introduction}

Almost fifty years ago Gel'fand and Tsetlin (GZ) discovered the 
explicit form of irreducible representations of the classical semisimple
Lie algebras such as $A_n,B_n$ and $D_n$ \cite{Tsetlin}. 
Their purely algebraic arguments were based on the 
possibility of the consequent embedding of one algebra into another. 
Under such kind of embedding the number of necessary additional ``quantum 
numbers'' is equal to the number of arising Cazimir operators.
   
   In the present paper we consider GZ results from another
point of view in order to connect them with some exactly integrable system
of finite difference equations or with its continuous limit which 
has a form of partial differential equations in the n-dimensional space.
The key point is that we propose the operator realization of
the generators of the simple roots $X^{\pm}_i$ and Cartan elements
$h_i$ of the $A_n$ algebra.
The selection rules used by us (the assumed form of operators) are
of course initiated by G-Z results.

As a consequence of the commutation relations between $3n$ elements of
the $A_n$ algebra ($2n$ generators of its simple roots $X^{\pm}_i$
and $n$ elements $h_i$ of its Cartan subalgebra)
\begin{equation}
  [X^+_i,X^-_j]=\delta_{i,j}h_i, \quad [h_i,X^{\pm}_j]=\pm k_{j,i} X^{\pm}_j,
\label{1}
\end{equation}
in some specific form of their realization
arises a discrete system of equations for $n$ unknown functions
depending on $n$ arguments. 
This system allows the exact integration.
Finally, the particular solution of this system 
under some additional requirements of irreducibility 
overgoes to the matrix elements the famous GZ paper.

It is more suitable to perform continuous limit not from the limiting 
procedure from discrete variables but by independently 
considering (\ref{1}) on the level of Poisson brackets.
Then we can use the Darboux theorem for resolving 
the arising functional group. (We use the old sense of this term, 
see, {\it e.g.,} the book of Eisenhart \cite{Eisenhart}) 
In this way it is possible to obtain the GZ expressions 
in the classical region. This method is much simpler 
technically  and may be used as additional hint 
for introducing and understanding the algebraic GZ formulae.     

Such general approach to the problem of the constructing the irreducible
representations of the Lie algebras was proposed many years ago in 
\cite{Malkin} but up to now has not been used (to the best of our
knowledge).

We begin our discussion in sections 2,3 with the simplest cases of $A_1$
and $A_2$ algebras, discribing in detail all necessary steps 
of calculations.
In section 4 we consider the general case of $A_n$ algebra.
In section 5 we summarize the results of the paper and discuss  
the perspectives of the further investigations.

\section{The case of $A_1 \simeq SU(2)$ and $U(2)$)}

Every section will be divided into two parts: the 
classical case (the functional group level) 
and proper algebraic construction (''quantum case''). 
As it was remarked in the introduction, 
the classical results may be used as  good hint in
further algebraic (quantum) calculations.

\subsection{The functional algebra case}

The functional $U(2)$ algebra contains four elements 
$X^{\pm},H,I$ connected among themselves by Poisson brackets:
\begin{equation}
 \{ I,X^{\pm}\}=\{I,H\}=0,\quad \{X^+,X^-\}=H,
 \quad \{H,X^{\pm}\}= \pm 2 X^{\pm} \label{2}
\end{equation}
In correspondence with the Darboux theorem \cite{Eisenhart} out of the 
elements of the $U(2)$ functional group (\ref{2}) it is possible to construct 
a pair of canonically conjugated variables $M,m,\; \{M,m\}=1$ 
and two cyclic variables  $I,L\equiv X^+X^-+{1\over 4} H^2$ 
which have zero Poisson brackets among themselves and with all the 
other elements of the $U(2)$ functional group (clearly, up to 
an arbitrary canonical transformation).
Let us choose $M=H$ and $m={1\over 4} \ln {X^+ \over X^-}+f(H)$. 
With the help of the Poisson brackets (\ref{2}) it is not difficult 
to see that thus constructed $M,m$ are really canonically 
conjugate variables. 
Resolving these relations with respect to the functional group elements 
leads to following realization of the functional group elements 
in terms of the canonical conjugated coordinate $m$ and momentum $M$
and the two cyclic momenta $L_1,L_2$:
\begin{eqnarray}
 X^+&=&{1\over 2}e^{2m} (L_1-M),\quad X^-={1\over 2}e^{-2m} (M-L_2),\quad
 H=M, \nonumber \\
 K^1I&=&L_1+L_2,\quad K^2={1\over 4}({L_1-L_2\over 2})^2.     \label{3}
\end{eqnarray}
By direct calculations it is not difficult to verify that (\ref{3})
is indeed a realization of the functional group (\ref{2}).
If we want to restrict ourselves with the case of $A_1$ algebra 
it is necessary to put $I=0$.

\subsection{Algebraic case}

As always, to pass from the classical expressions to the quantum ones
it is necessary to order in some way the operators involved 
and replace the Poisson brackets by commutators. 
The equations (\ref{3}) give us a hint about the very tempting  
possibility to rewrite them as
\begin{eqnarray}
  X^+&=&{1\over 2}e^m (L_1-M) e^m,\quad X^-={1\over 2}e^{-m} (M-L_2)
    e^{-m},\nonumber \\
  H&=&M-{L_1+L_2\over 2},\quad Q=L_1+L_2,\quad L={1\over 2}({L_1-L_2\over 2}^2
  -1)
\label{4}
\end{eqnarray}
and consider now $M,m$ as generators of the Heisenberg algebra
($[M,m]=1,\;[M,1]=0,\;[m,1]=0$, and $L_1,L_2$ commute with
all of the generators involved in (\ref{4})). 
 Keeping in mind the following operator relation for the Heisenberg algebra,
$e^{\pm x} p e^{\mp x}=p\mp 1$, we conclude that the generators,
defined in (\ref{4}) satisfy the commutation relations (\ref{2}) 
of $U(2)$ algebra (of course with square brackets instead of 
the curly ones).  

Two Cazimir operators under realization (\ref{4}) take on the constant 
values
\begin{equation}
 K^{(1}=L_1+L_2, \quad K^{(2}=X^+X^-+ X^-X^+ +{1\over 2}\,H^2=({L_1-L_2)\over 
 2}^2 -1),\label{5}
\end{equation}
which proves the irreducibility of the constructed representation.

\section{$A_2 \simeq SU(3)$ and $U(3)$ cases}

In this case the problem consists in resolving the system of commutation
relations
\begin{eqnarray}
 \left[X^+_1,X^-_1\right]=h_1,\quad \left[X^+_1,X^-_2\right]=0\quad 
\left[h_1,X^{\pm}_1\right]=\pm 2X^{\pm}_1,
   \quad \left[h_1,X^{\pm}_2\right]=\mp X^{\pm} \nonumber \\
 \left[X^+_1,X^-_2\right]=0,\quad \left[X^+_2,X^-_2\right]=h_2\quad 
\left[h_2,X^{\pm}_2\right]=\pm 2X^{\pm}_2,
\quad \left[h_2,X^{\pm}_1\right]=\mp X^{\pm}_1 . \label{6}
\end{eqnarray} 
Algebra representation theory insures us that this is equivalent to the 
construction of some representation (possibly a reducible one) of the 
$A_2$ algebra.

The "selection rules" of GZ paper allow us to try to find a solution of
this problem in the following form:
\begin{eqnarray}
 X^+_1 &=&{1\over 2}e^m(L_1-M) e^m,\quad X^+_2=e^{l_1}f^1e^{l_1}+
   e^{l_2} f^2 e^{l_2},\quad \nonumber \\
 h_1 &=& M-{L_1+L_2\over 2} \nonumber \\
  X^-_1 &=& {1\over 2}e^{-m}(M-L_2) e^{-m},\quad X^-_2=e^{-l_1} \bar f^1
e^{-l_1}+e^{-l_2} \bar f^2 e^{-l_2},\nonumber \\
 h_2 &=& -{M\over 2}+L_1+L_2- {N_1+N_2+N_3\over 2},     \label{7}
\end{eqnarray}     
where all "structural" functions $f^{1,2},\bar f^{1,2}$ depend only on
momentum (capital letters) variables. We intentionally preserve the order of
operators to avoid rewriting the same formulae for several 
times.

\subsection{Classical case}

In this case it is necessary to understand all of the above relations   
at the functional group level. The commutators have to be replaced by  
the Poisson brackets understood as usually:
$$
\{A,B\}=\sum^3_1 \left(\frac {\partial A}{\partial p_i}
  \frac {\partial B}{\partial x_i}-
  \frac {\partial A}{\partial x_i}\frac {\partial B}{\partial
   p_i}\right),
   \quad x_i=(m,l_1,l_2),\quad p_i=(M,L_1,L_2)
$$
Now all of the objects are commutative and the order of the factors
in (\ref{7}) is unimportant.
As a consequence of the vanishing Poisson brackets
$\{X^+_1,X^-_2\}=\{X^+_2, X^-_1\}=0$,
we obtain the explicit dependence of structural functions on momentum
$M$. Namely,
\begin{eqnarray}
  X^+_1 &=& {1\over 2}e^m(L_1-M) e^m,\quad X^+_2=e^{l_1}f^1e^{l_1}+
      e^{l_2}(M-L_2) f^2 e^{l_2},\nonumber\\
  X^+_{12} &\equiv & \left[X^+_1,X^+_2\right]=e^m \{e^{l_1}f^1e^{l_1}-
     e^{l_2}(L_1-M) f^2 e^{l_2}\}e^m  \nonumber\\
 h_1 &=& M-{L_1+L_2\over 2},    \quad
    h_2=-{M\over 2}+L_1+L_2- {N_1+N_2+N_3\over 2}  \label{9}\\
 X^-_{12} &\equiv & \left[X^-_2,X^-_1\right]=
    e^{-m} \{e^{-l_1} (L_1-M)\bar f^1e^{-l_1}-
     e^{-l_2}\bar f^2 e^{-l_2}\}e^{-m} \nonumber\\
 X^-_1 &=& {1\over 2}e^{-m}(M-L_2) e^{-m},\quad
     X^-_2=e^{-l_1} (L_1-M)\bar f^1
     e^{-l_1}+e^{-l_2} \bar f^2 e^{-l_2} . 
\end{eqnarray}
Keeping in mind the above arguments, we have preserved the order of factors 
essential for the consideration in the next subsection. 
Moreover, we have presented also the explicit form of 
the generators of the composed roots $X^{\pm}_{12}$,
which will be necessary below for constructing of 
the Cazimir operators.
  The additional constant in $h_2$ will be explained a little bit later.
In the case of the $A_2$ algebra this constant is equal to zero.

The remaining Poisson bracket $\{X^+_2,X^-_2\}=h_2$, unused up to now,
has as its consequence the system of equations for structural functions. 
Technically, all these operations are simple and we represent only the rezult.  
\begin{equation}
  (f^1\bar f^1)_{L_1}-(f^2\bar f^2)_{L_2}=-{1\over 4},\quad -(L_1f^1\bar f^1)_
  {L_1}+(L_2f^2\bar f^2)_{L_2}={L_1+L_2\over 2}-{N_1+N_2+N_3\over 4}
  \label{A9}
\end{equation}
\begin{equation}
  f^2_{L_1}\bar f^1+f^2\bar f^1_{L_2}=0,\quad f^1_{L_2}\bar f^2+f^1\bar f^2_
  {L_1}=0 \label{B9} .
\end{equation}
By shifts of independent variables $L_1\to L_1+ (N_1+N_2+N_3)/3$ and 
$L_2\to L_2+(N_1+N_2+N_3)/3$ the
constant term from the second equation (\ref{A9}) may be taken away.
This corresponds to the transition from the algebra $U(3)$ to $SU(3)$.

{}From now on we can divide our problem into two parts and formulate 
it more precisely.
First of all, we want to prove that the linear system of equations (\ref
{A9}) is exactly integrable and to give its general solution.
In the case of arbitrary $A_n$ algebra it will be the linear system 
of equations 
in partial derivatives for $n$ unknown function in the n-dimensional space.
Secondly, we can solve  this  system  under  additional 
restrictions, which follow from (\ref{B9}), and in this way to obtain once 
again the GZ formulae for matrix elements of irreducible representations of
the $A_n$ algebra.

In this section we present the solution of the second part of the problem.
The functional group possesses two cyclic variables (Cazimir operators
on the algebra representation level), which may be constructed as
the traces of the second and third degree of the following matrix
$$
K=\pmatrix{ h^1   & X^-_1   & X^-_{12} \cr
            X^+_1 & h^2-h^1 & X^-_2 \cr
         X^+_{12} & X^+_2   & -h^2   }
$$
where $h^1={(2h_1+h_2)\over 3} ={M\over 2}, $ and
$ h^2={(h_1+2h_2)\over 3}={(L_1+L_2)\over 2}$.
Explicit forms of Cazimir operators of the second- and the third-order
are as follows:
$$
 K^{(2}=X^+_{12} X^-_{12}+X^+_1X^-_1+X^+_2X^-_2+(h^1)^2-h^1h^2+(h^2)^2=
 (L_1-L_2)(f^1\bar f^1+f^2\bar f^2)+ {1\over 4}(L_1^2+L_1L_2+L_2^2),
$$
$$
K^{(3}=X^+_{12}X^-_1X^-_2+ X^-_{12}X^+_1X^+_2-h^1X^+_2X^-_2+h^2X^+_1X^-_1+
(h^1-h^2)X^+_{12} X^-_{12}+h^1h^2(h^1+h^2)=
$$
$$
{1\over 2}\{(L_1-L_2)(L_1f^1\bar f^1+L_2f^2\bar f^2)+ {1\over 4}
L_1L_2(L_1+L_2).
$$

Equating the Cazimir operators to constant values, we obtain additional
system of linear algebraic equations for the functions $x\equiv f^1\bar f^1, 
y\equiv f^2\bar f^2$, which is consistent with (\ref{A9}), (\ref{B9}):
$$
  x+y={\sigma_2-(L_1^2+L_1L_2+L_2^2)\over 4(L_1-L_2)^2},\quad
  L_2x+L_1y={\sigma_3-L_1L_2(L_1+L_2)\over 4(L_1-L_2)^2}
$$
The solution of the last system is the given as follows,  
\begin{equation}
 f^1\bar f^1={P_3(L_1)\over 4(L_1-L_2)^2},\quad
 f^2\bar f^2={-P_3(L_2)\over 4(L_1-L_2)^2} , \label{10}
\end{equation}
where $P_3(z)=(z-N_1)(z-N_2)(z-N_3)$ and $N_1+N_2+N_3=0$ in
the case of $A_2$ algebra.
We emphasize that the solution (\ref{10}) is only a particular
solution of the system (\ref{A9}), but not of the general one.

\subsection{General solution of the linear system}

To stress the symmetry properties of (\ref{A9}), we rewrite it
in the variables
$X^1\equiv f^1\bar f^1,X^2\equiv -f^2\bar f^2$, $x_1\equiv L_1,
x_2\equiv L_2$, keeping only the homogeneous part of it:
\begin{equation}
X^1_{x_1}+X^2_{x_2}=0,\quad (x_1X^1)_{x_1}+(x_2X^2)_{x_2}=0 \label{11}.
\end{equation}
The system (\ref{11}) is a particular case of the following system for $n$
unknown functions $X^i$ in $n$-dimensional coordinate space $x_i$:
\begin{equation}
 \sum^n_{i=1} (x^k_i X^i)_{x_i}=0 ,\quad k=0,1,2,...n-1 \label{12}
\end{equation}
the particular case (\ref{11}) corresponds to the choice $n=2$ in (\ref{12}).

Now we would like to demonstrate the direct way for obtaining the general
solution of (\ref{11}). Multiplying the first equation of the system
(\ref{11}) correspondingly by $x_1,x_2$ and subtracting it from the second 
equation we obtain:
$$
((x_2-x_1)X^2)_{x_2}+X^1=0,\quad ((x_1-x_2)X^1)_{x_1}+X^2=0.
$$
Introducing new unknown functions $Y^{1,2}\equiv (x_1-x_2)^2X^{1,2}$, we
transform the system (\ref{11}) to the  form
$$
(x_2-x_1)(Y^2)_{x_2}+Y^1-Y^2=0,\quad (x_1-x_2)(Y^1)_{x_1}+Y^2-Y^1=0 ,
$$
from which we conclude that $(Y^1)_{x_1}=(Y^2)_{x_2}$. 
At last, differentiating  the first equation with respect to 
$x_1$ and the second one with respect to  $x_2$,
we present each of them in the integrable form:
\begin{equation}
\frac{\partial \ln Y^2_{x_1}}{\partial x_2}={1\over x_2-x_1}\quad
\frac{\partial \ln Y^1_{x_2}}{\partial x_1}={1\over x_1-x_2} \label{13}
\end{equation}
Finally, the general solution of the system (\ref{11}) takes the form:
$$
(x_1-x_2)^2X^1=\Phi(x_2)_{x_2}(x_1-x_2)+\Theta(x_1)+\Phi(x_2),
$$
\begin{equation}
(x_1-x_2)^2X^2=\Theta(x_1)_{x_1}(x_2-x_1)+\Phi(x_2)+\Theta(x_1) \label{14}
\end{equation}
where $\Theta(x_1),\Phi(x_2)$ are arbitrary functions of a single argument.

\subsection{Algebraic cases $A_2\simeq SU(3)$ and $U(3)$}

In this section we will consider a realization of (\ref{9}), where
the pairs $M,m$, $L_1,l_1$, $L_2,l_2$ are the elements of the 
three independent mutually commutative Heisenberg algebras 
$\left[M,m\right]=1,\;[L_1,l_1]=1,\;[L_2,l_2]=1$.
The involved structural functions are supposed to depend only on
the momentum operators $M,L_1,L_2,N_1,N_2,N_3$.

Explicit dependence of the structural functions upon the momentum $M$ arises
immediately, if we take into account commutation relations $[X^+_1,X^-_2]=
[X^+_2,X^-_1]=0$.

Commutation relation $[X^+_2,X^-_2]=h_2$, after using
the well-known from the theory of the Heisenberg algebra relations
$\exp(\pm x) \,p \exp (\mp x) = p\mp 1$
is equivalent to the system of equations in finite differences, 
which will be convenient to write in the following notation:
$$
F^{(1}_{\pm}=f^{(1}(L_1\pm 1,L_2)\bar f^{(1}(L_1\pm 1,L_2) , \quad
F^{(2}_{\pm}=f^{(2}(L_1,L_2\pm 1)\bar f^{(2}(L_1,L_2\pm 1)
$$
\begin{eqnarray}
   F^{(1}_+-F^{(1}_-+F^{(2}_+-F^{(2}_-  &=&-{1\over 2},\nonumber \\
      -(L_1+1)F^{(1}_++(L_1-1)F^{(1}_-  &-&(L_2-1)F^{(2}_-+(L_2+1)F^{(2}_+ 
     \nonumber \\
                                        &=& L_1+L_2-{N_1+N_2+N_3\over 2}             
   \label{A14}  \\
  f^{(2}(L_1- 1,L_2)\bar f^{(1}(L_1,L_2-1) &=&f^{(2}(L_1+ 1,L_2)\bar 
    f^{(1}(L_1, L_2+1),  \nonumber \\
  f^{(1}(L_1,L_2-1)\bar f^{(2}(L_1-1,L_2)  &=&f^{(1}(L_1,L_2+1)
    \bar f^{(2}(L_1+1,L_2). \label{B14}
\end{eqnarray}
In the continuous limit (\ref{A14}) coincides with (\ref{A9}), and
(\ref{B14}) coincides with (\ref{B9}).

The Cazimir operators may be constructed as traces of the first,
second and third degrees of the $K$ matrix as in the previous subsection.
But in this case it is necessary to take into account 
noncommutativity of operators involved. In this way we obtain 
their explicit expressions:
\begin{eqnarray}
  K^{(1} &=& {N_1+N_2+N_3\over 2} \nonumber\\
  K^{(2} &=& X^+_{12} X^-_{12}+X^-_{12} X^+_{12}+
     X^+_1X^-_1+X^-_1X^+_1+X^+_2X^-_2+
     X^-_2X^+_2+2((h^1)^2-h^1h^2+(h^2)^2) \nonumber\\
  K^{(3} &=& X^-_2X^+_{12}X^-_1+ X^+_1X^-_{12}X^+_2-
     {1\over 2}h^1\{X^+_2X^-_2\}+
     {1\over 2}h^2\{X^+_1X^-_1\}+
     {1\over 2}(h^1-h^2-2)\{X^+_{12} X^-_{12}\}+ \nonumber\\
   & & h^1h^2(h^1+h^2)+(h^1+h^2)-((h^1)^2-h^2h^1+(h^2)^2)
\end{eqnarray}
Equating Cazimir's operators to constant values, we obtain the additional
and consistent with (\ref{A14}),(\ref{B14}) equations:
\begin{eqnarray}
 && F^{(1}_++F^{(1}_++F^{(2}_++F^{(2}_-={\sigma_2-(L_1^2+L_1L_2+L_2^2)\over 2
    (L_1-L_2)} \nonumber\\
 && \left[{(L_2-1)(L_1+L_2+4)\over 4} -(L_1+1)(L_2+1)\right]F^{(1}_+ +
     {(L_2+1)(L_1+L_2+4) \over 4} F^{(1}_-\nonumber\\
 &&     -{(L_1+1)(L_1+L_2+4)\over 4}F^{(2}_-
    +\left[ (L_2+1)(L_1+1)-{(L_1-1)(L_1+L_2+4)\over 4}\right]
      F^{(2}_+ =\nonumber\\
 &&   \sigma_3+{1\over 8}L_1L_2(L_1+L_2)+
         {5\over 8}(L_1+L_2)+{1\over 4}(L_1+L_2)^2      \label{nomer}
\end{eqnarray}
In connection with the last system the following comments
will be appropriate.
(\ref{A14}) is the system of the two equations for two unknown functions
$F^{(1,2}$ with shifted $\pm 1$ arguments. In this sense it is a
closed one. 
(\ref{B14}) and (\ref{nomer}) may be considered as additional
to (\ref{A14}) conditions, by which are necessary to choose from the general
solution only those that correspond to the realization of the irreducible
(Cazimir operators are fixed) representations  of the $A_2$ algebra.

As in the previous section, here the two problem arise: 
to find a general solution of the system (\ref{A14})
and ,as the second step, to satisfy additional
conditions which follow from (\ref{B14}) and (\ref{nomer}).

Here we present a solution of the second part of the above formulated problem.
Combining the first
equation of ({\ref{A14}) with the first equation from (\ref{nomer}), we 
immediately obtain:
$$
F^{(1}_++F^{(2}_-=-{1\over 4}+{\sigma_2-(L_1^2+L_1L_2+L_2^2)\over 4(L_1-L_2)},
\quad
F^{(1}_++F^{(2}_+={1\over 4}+{\sigma_2-(L_1^2+L_1L_2+L_2^2)\over 4(L_1-L_2)}
$$
For unknown functions $,v\equiv (L_1-L_2-2)(F^{(1}_--F^{(2}_+),u\equiv
(L_1-L_2+2)(F^{(1}_+-F^{(2}_-)$ we obtain the linear
system from the remaining equations
$$
v-u={3\over 2}(L_1+L_2)
$$
$$
(L_1-L_2+2)v+(L_1-L_2-2)u={(L_1+L_2+4)(\sigma_2-(L_1^2+L_1L_2+L_2^2))\over 4}+
$$
$$
L_1L_2(L_1+L_2)+2(L_1^2+L_1L_2+L_2^2)+2(L_1+L_2)+\sigma_3
$$
with the explicit solution:
\begin{equation}
4(L_1-L_2-1)(L_1-L_2+1)F^{(1}=P_3(L_1),\quad 4(L_1-L_2-1)(L_1-L_2+1)F^{(2}=
P_3(L_2) \label{19}
\end{equation}
where $P_3(z)=(z-N_1)(z-N_2)(z-N_3)$.

The square roots of $F^{(1},F^{(2}$ are exactly the matrix elements
of Gel'fand-Tsetlin realization of $U(3)$ algebra.

\section{General case of arbitrary $n$}

\subsection{The algebra representation level}

Let us assume that the generators of the simple roots and Cartan 
elements of elements of $U(n+1)$ algebra may be represented in the form
\begin{equation}
X^+_s=\sum_{k=1}^s e^{ l^s_k} g^s_k e{l^s_k},\quad X^-_s=\sum_{k=1}^s
e^{-l^s_k} \bar g^s_k e^{-l^s_k} \label{20}
\end{equation}
$$
h_s=-{1\over 2}\sum_{r=1}^{s-1} L^{s-1}_r+\sum_{k=1}^s L^s_k-{1\over 2}
\sum_{l=1}^{s+1} L^{s+1}_l,\quad 1\leq s \leq n
$$
where non zero commutators of operators involved are only those
$$
 [L^s_k,l^t_l]=\delta_{st} \delta_{kl}I
$$
and structural functions are the functions only of the following arguments: 
$g^s_k\equiv g^s_k(L^{s+1},L^s,L^{s-1}), $
$ \bar g^s_k\equiv \bar g^s_k(L^{s+1},L^s,L^{s-1})$. The reader may identify 
without any difficalties $L^1$ with $M$, $L^2$ with $L_1,$ $L_2$ and at
last $L^3$ with $N_1,N_2,N_3$.
We will assume also that structural functions may be represented in 
factorizable form
$$
  g^s_k\equiv F^s_k(L^{s+1};L^s) f^s_k(L^s;L^{s-1})\, \quad 
 \bar g^s_k\equiv \bar F^s_k(L^{s+1};L^s) \bar f^s_k(L^s;L^{s-1})\,,
$$
and 
$$
(F^{s-1}_k)^2=(\bar F^{s-1}_k)^2={\prod_{r=1}^s (L^{s-1}_k-L^s_r)\over
\Phi (L^{s-1})}
$$
where function $\Phi$ is translation invariant with respect to the shift of
all of its arguments $L^{s-1}_k$.

The last proposition we will prove by induction.

It is obvious that under such kind of restrictions commutation relation
between generators of Cartan subalgebra $h_l$ and generators of the simple
roots $ X^{\pm}_k $ are correctly satisfied. It is also clear that 
the generators $ X^{\pm}_s $ commute with all generators $X^{\mp}_k$ with
$ k\leq (s-2) $ because they act on essentially different arguments.
And finally, commutation relations
$$
  [X^{\pm}_s,X^{\mp}_{s-1}]=0
$$
allow us to reconstruct in explicit form the dependence of 
structural functions
$f^s_k,\bar f^s_k$ on arguments $L^{s-1}_k$.

As a direct consequence of the last commutation relations we obtain:
$$
f^s_k(L^s;..,L^{s-1}_r-1,..)\bar F^{s-1}_r(..,L^s_k-1,..;L^{s-1})=
f^s_k(L^s;..,L^{s-1}_r+1,..)\bar F^{s-1}_r(..,L^s_k+1,..;L^{s-1})
$$
$$
\bar f^s_k(L^s;..,L^{s-1}_r-1,..) F^{s-1}_r(..,L^s_k-1,..;L^{s-1})=
\bar f^s_k(L^s;..,L^{s-1}_r+1,..) F^{s-1}_r(..,L^s_k+1,..;L^{s-1})
$$

The last relations must be satisfied for all possible
values of the indices $k$ and $r$. Keeping in mind the explicit form of the
structural functions $F^{s-1}_k=\bar F^{s-1}_k$, proposed above, we can to
resolve the last equations in the form :
\begin{equation}
g^s_k =F^s_k(L^{s+1};L^s)\sqrt{\prod^{s-1}_{r=1} (L^s_k-L^{s-1}_r)},\quad
\bar g^s_k=\bar F^s_k(L^{s+1};L^s)\sqrt{\prod^{s-1}_{r=1} (L^s_k-L^{s-1}_r)}
\label{21}
\end{equation}

Up to now we have not satisfied the only commutation relation
\begin{equation}
 [X^+_s,X^-_s]=h_s \label{22}
\end{equation}

This equation contains a "diagonal" part (which does not contain 
the coordinates of Heisenberg subalgebras $l_i$) and non-diagonal one 
(in above sense).
It is suitable to write the emereged equation for the diagonal part in
the following notations:
$$
X^{\pm s}_k=F^s_k(L^{s+1};..,L^s_k \pm 1,..)\bar F^s_k(L^{s+1};..,L^s_k \pm
1...)
$$
\begin{equation}
\sum_{k=1}^s \prod^{s-1}_{r=1} (L^s_k+1-L^{s-1}_r) X^{+s}_k-\sum_{k=1}^s
\prod^{s-1}_{r=1} (L^s_k-1-L^{s-1}_r) X^{(-s}_k=   \label{23}
\end{equation}
$$
-{1\over 2}\sum_{r=1}^{s-1} L^{s-1}_r+\sum_{k=1}^s L^s_k-{1\over 2}
\sum_{j=1}^{s+1} L^{s+1}_j
$$
The left-hand side of this equation may be represented as linear combination
of $s-1$ symmetrical functions, constructed out of $(s-1)$ 
momentum variables $L^{s-1}_r$. 
The right-hand side of (\ref{23}) contains only terms of the first 
and zero degree with respect to such functions. Thus, as a corollary of 
(\ref{23}), we obtain the system of $s$ equations in finite differences
for determining $s$ unknown functions $X^s_k$:
$$
\sum_{k=1}^s [(L^s_k+1)^{s-1} X^{+s}_k-(L^s_k-1)^{s-1} X^{-s}_k]=
\sum_{k=1}^s L^s_k-{1\over 2}\sum_{j=1}^{s+1} L^{s+1}_j
$$
\begin{equation}
\sum_{k=1}^s [(L^s_k+1)^{s-2} X^{+s}_k-(L^s_k-1)^{s-2} X^{-s}_k]={1\over 2}
\label{24}
\end{equation}                                                              
$$
\sum_{k=1}^s [(L^s_k+1)^i X^{+s}_k-(L^s_k-1)^i X^{-s}_k ]=0\quad
0\leq i \leq (s-3)
$$

The zero value of "non-diagonal" part of (\ref{22}) is equivalent to 
additional conditions which structural functions must satisfy:
$$
F^s_k(L^{s+1};..,L^s_j-1,..) \bar F^s_j(L^{s+1};..,L^s_k-1,..)=
F^s_k(L^{s+1};..,L^s_j+1,..)\bar F^s_j(L^{s+1};..,L^s_k+1,..)
$$
\begin{equation}
{}\label{25}
\end{equation}
$$
\bar F^s_k(L^{s+1};..,L^s_j-1,..) F^s_j(L^{s+1};..,L^s_k-1,..)=
\bar F^s_k(L^{s+1};..,L^s_j+1,..) F^s_j(L^{s+1};..,L^s_k+1,..)
$$
{}From (\ref{25}) we see that the solution $F^s_j=\bar F^s_j$
is possible and the functions $X^s_k$ (as the solutions of (\ref{23})) 
must satisfy additional conditions:
\begin{equation}
X^s_k(L^{s+1};..,L^s_j-1,..) X^s_j(L^{s+1};..,L^s_k-1,..)=
X^s_k(L^{s+1};..,L^s_j+1,..) X^s_j(L^{s+1};..,L^s_k+1,..)\label{26}
\end{equation}
Putting $s=2$, the reader can easily obtain 
from general equations of the present subsection all results of 
the previous section,  in particular, equations (\ref{A14}). 
At this point we interrupt our consideration for a moment in order to 
represent a general solution of the continuous 
version of equations (\ref{24}). 

\subsection{General  solution  of  the  linear   system   in   the
continuous limit}

The continuous limit of the homogeneous part of the system (\ref{24}) in
variables $X^i\equiv F^n_i,x_i\equiv L^n_i$ has the form
\begin{equation}
 \sum_{i=1}^n (x_i^k X^i)_{x_i}=0 ,\quad k=0,1,....,n-1  \label{26a}
\end{equation}

We do not know the simple regular methods of the direct resolution of
(\ref{26a}). The way of solving of this problem known to us for 
the particular case $n=3$ is given in Appendix. Here we represent 
the final result and the proof of its validity.

General solution of the system (\ref{26a}) is given by the formula
\begin{equation}
  X^s=\left({\Theta\over \prod_{k=1}^{'n} (x_i-x_k)}
   \right)_{x_s..x_{s-1}x_{s+1}..x_n}
\label{27}
\end{equation}
where the function $\Theta$ satisfies the equation $ \Theta_{x_1...x_n}=0$
(here the differentiation is performed with respect to 
all independent coordinates of the problem!).

The following result from the theory of symmetrical functions will be
necessary for us to prove that (\ref{27}) is indeed a solution of 
(\ref{26a}):
$$
\sum_{i=1}^n {x_i^r\over \prod_{k=1}^{'n} (x_i-x_k)}=(0,1,S^{r-n+1}).
$$
In the parenthesis in the right-hand side different possibilities are 
given. 
The first one ($0$) takes place if $r$ is strictly less then $n-1$;
the second one occurs if $r=n-1$ and the third possibility corresponds to
the case when $r$ is strictly more the $n-1$ and $S^k$ is some symmetrical 
function of k-th degree. 

It is possible to understand the above proposition without any calculations. 
Indeed, the written sum is a symmetrical function. 
After reducing it to the common denominator we obtain the ratio of two 
$n$-dimensional  polynomial   functions,   one   of   which   (the 
denominator) 
is exactly the Wandermond determinant (an only function which is 
antisymmetrical with respect to permutation of each pair
of coordinates). Hence the numerator must also  be  antisymmetrical 
polynomial (since the ratio is symmetric!). This is impossible if $r$ is less
then $n-1$ (the degree of the numerator in this case is less than the
degree of the Wandermond determinant).

After these comments let us consider an arbitrary equation from the 
system (\ref{27}). We have consequently 
$$
\sum_{s=1}^n \left(x_s^r \left({\Theta\over \prod_{k=1}^{'n} (x_s-x_k)}
\right)_{x_1..x_{s-1} x_{s+1}..x_n}\right)_{x_s}
=\left(\sum_{s=1}^n {x_s^r\over \prod_{k=1}^{'n} (x_s-x_k)}
\Theta\right)_{x_1...x_n}=0 .
$$
In the case when $r$ is strictly less then $n-1$ the sum by itself
equals to zero; in the case when $r=(n-1)$ the sum is equal to unity 
but the condition for the function $\Theta$ equates to zero the 
corresponding equation of the system.

Thus, our proposition is proved and (\ref{27}) is indeed 
a general solution of the linear system (\ref{26a}).

\subsection{Continuation of the discussion}

Now we can present the solution of inhomogeneous system (\ref{25})
together with the additional conditions (\ref{26}). We begin
with the continuous case, using the notations of the previous
subsection. In this case system (\ref{25}) takes the form:
\begin{equation}
 2\sum_{i=1}^n (x_i^{n-1} X^i)_{x_i}=-{1\over 2}\sum_{j=1}^{n+1} y_j+
 \sum_{i=1}^n x_i\quad 2\sum_{i=1}^n (x_i^{n-2} X^i)_{x_i}={1\over 2}
 \label{I}
\end{equation}
$$
\sum_{i=1}^n (x_i^r X^i)_{x_i}=0, \quad 0 \leq r \leq n-3
$$

Keeping in mind the solution for the particular case $n=2$ (\ref{19}) 
we will try to find a solution of the last system in the form: 
\begin{equation}
X^i={1\over 4}({P_{n+1}(x_i)\over \prod_{k=1}^{'n}(x_i-x_k)})_{x_1..x_{i-1}
x_{i+1}..x_n}\equiv{1\over 4}{P_{n+1}(x_i)\over \prod_{k=1}^{'n}(x_i-x_k)^2}
\label{II}
\end{equation}
where $P_{n+1}(x_i)=\prod_{j=1}^{n+1}(x_i-y_j)$. Substituting (\ref{II})
into each equation (\ref{I}) we consequently check all equations:
$$
\sum_{i=1}^n (x_i^r X^i)_{x_i}={1\over 4}({x^rP_{n+1}(x_i)\over \prod_{k=1}^
{'n}(x_i-x_k)})_{x_1..x_n}, 
$$
where in the last expressio it is necessary to perform the differentiation
on all coordinates.

In connection with the facts from the theory of the symmetrical functions,
mentioned in previous subsection, the sum under the symbol of differentiation 
in the case $(0\leq r\leq (n-3))$ is nothing else but the linear combination 
of the symmetrical functions with the degree not greater than $n-1$. After
differentiating with respect to all coordinates such functions vanish. 
By the same reason only the term with the highest power 
of polynomial gives contribution different from zero in the case when 
($r=n-2$) and the terms $x_i^{n+1}-(\sum_{j=1}^{n+1} y_j) x_i^n$ in the case 
when $r=n-1$. 
Numerical values for corresponding sums, which can be calculated without 
big difficulties, shows that the first pair of equations (\ref{I}) is also
satisfied. It is not difficult to check that the solution (\ref{II}) satisfies 
additional conditions (\ref{26}).

In the algebraic case situation is very similar. To show this let us 
introduce the operation of the discrete differentiation, defined as 
$$
\Delta_i f(x_1,...x_n)={f(...x_i+1,...)-f(....x_i-1,....)\over 2}.
$$
Operations $\Delta_i$ and $\Delta_j$ are obviously mutually commutative 
and satisfy the linear condition in the following sense:
$$
\Delta_i (f^1+f^2)=\Delta_i f^1+\Delta_i f^2,\quad \Delta_i c f=c\Delta_i f, 
$$
if $c$ is some function independent of the $x_i$ coordinate. 

In this notations general solution of homogeneous system (\ref{24}) has the 
form
\begin{equation}
X^i=(\Delta_1 ...\Delta_{i-1}\Delta_{i+1}...\Delta_n){\Theta \over \prod_{k=1}
^{'n} (L^n_i-L^n_k)}, \label{III}
\end{equation}
where the function $\Theta$ is an arbitrary solution of the equation in
finite differences 
$$
(\Delta_1 .....\Delta_n)\Theta=0 .
$$ 
Solution of inhomogeneous system (\ref{24}) satisfying additional conditions
(\ref{26}) is given by
\begin{equation}
F^i=(\Delta_1 ...\Delta_{i-1}\Delta_{i+1}...\Delta_n){P_{n+1}(L^n_i)\over 
\prod_{k=1}^{'n} (L^n_i-L^n_k)}\,,  
\quad 
   P_{n+1}(x) = \prod_{j=1}^{n+1} \left( x-L^{n+1}_j\right)\,.
 \label{IIII}
\end{equation}

We omit the proofs of the last propositions because it does not change
in essential points in a compare with the continuous case. This fact is 
extremely interesting and remarkable and for it will take some time comprehend 
it.

\section{Outlook and possible perspectives}

In the present paper we have presented some new (unknown before or may be
forgotten at present moment) linear integrable system in the space of
arbitrary $n$ dimensions. This system is invariant with respect to
transformations from the whole group of permutations of $n$ symbols.
Particular solutions of this system are in the deep connection with
the algebra $A_n$ and matrix elements of its irreducible representations,
discovered many years ago by Gel'fand and Tsetlin \cite{Tsetlin}. 
This fact allows us to reproduce GZ result by a new method and 
moreover, to make them more understandable. 
Remarkably, under our formulation of the problem it was 
sufficient to perform calculations only on the functional group 
level. But give any explanation or comment of this fact we are now
not ready.

By some reasons (which are not clear for us at this moment) 
the quasiclassical approach gives functionally correct quantum result.
We hope that the method proposed in this paper can be generalized 
to the case of arbitrary semisimple algebra, 
relating the used parameterization with the properties of the Weyl
discrete group of corresponding semisimple algebra.

We hope also that by a method of the present paper it will be possible 
to find new realizations of quantum and deformed algebras.

Author is indebted to the
Instituto de Investigaciones en Matem\'aticas
  Aplicadas y en Sistemas, UNAM, Mexico
for beautiful conditions for his work.

Permanent company and discussions with N.M.\ Atakishiyev, S.M.\ Chumakov,
K.B.\ Wolf and P.\ Winternitz allow the author to finish this paper in the
shortest time.

\section{Appendix}

The initial system in the case $n=3$ has the form:
$$
X^1_{x_1}+X^2_{x_2}+X^3_{x_3}=0
$$
\begin{equation}
(x_1 X^1)_{x_1}+(x_i^k X^i)_{x_i}+(x_3 X^3)_{x_3}=0 \label{A1}
\end{equation}
$$
(x_1^2 X^1)_{x_1}+(x_2^2 X^2)_{x_2}+(x_3^2 X^3)_{x_3}=0,
$$
Multiplying the first equation by $x_2x_3$, the second one by $-(x_2+x_3$
and summing them with the third one we obtain:
$$
((x_3-x_2)(x_3-x_1)X^3)_{x_3}+{(x_1-x_2)(x_1-x_3)X^1\over (x_1-x_3)}+
{(x_2-x_1)(x_2-x_3)X^2\over (x_2-x_3)}=0
$$
Let us introduce the new unknown functions $\bar X^i\equiv \prod^{'n}_{k=1}
(x_i-x_k) X^i$ and rewrite the last equation (together with those which arise 
from it after obvious permutations of the indexes), we come to a system:
$$
\bar X^1_{x_1}+{\bar X^2\over (x_2-x_1)}+{\bar X^3\over (x_3-x_1)}=0
$$
\begin{equation}
\bar X^2_{x_2}+{\bar X^1\over (x_1-x_2)}+{\bar X^3\over (x_3-x_2)}=0
\label{A2}
\end{equation}
$$
\bar X^3_{x_3}+{\bar X^1\over (x_1-x_3)}+{\bar X^2\over (x_2-x_3)}=0
$$
The next transformation $\tilde X^i=\equiv \prod^{`n}_{k=1}(x_i-x_k) 
\bar X^i$ would be possible to realize independently on the first step, but 
the form of the system (\ref{A2}) will be necessary for further consideration 
and for this reason we represent it above. In the new variables we obtain
$$
(x_1-x_2)(\tilde X^2_{x_2}+\tilde X^1_{x_1})+2(\tilde X^2-\tilde X^1)=0
$$
\begin{equation}
(x_2-x_3)(\tilde X^3_{x_3}+\tilde X^2_{x_2})+2(\tilde X^3-\tilde X^2)=0
\label{A3}
\end{equation}
$$
(x_1-x_2)(\tilde X^1_{x_1}+\tilde X^3_{x_3})+2(\tilde X^1-\tilde X^3)=0
$$
After excluding unknown $\tilde X^3$ from the two last equations and
introducing $v\equiv \tilde X^2_{x_3}$ and $u\equiv \tilde X^1_{x_3}$ we
obtain system of two equations
$$
v_{x_2}-u_{x_1}-{2v\over x_2-x_3}-{2u\over x_3-x_1}=0,\quad
v_{x_2}+u_{x_1}+{2v\over x_1-x_2}-{2u\over x_1-x_2}=0
$$
Finally excluding unknown $u$ we obtain an equation for $v$ in
the integrable form:
$$
v_{x_1x_2}=v_{x_1}({1\over x_2-x_1}+{1\over x_2-x_3})
$$
Returning back, after some algebraic manipulations we obtain general solution
of the initial system in the form:
$$
\bar X^1=\Theta_{x_2x_3}+{\Theta_{x_3}\over x_1-x_2} +{\Theta_{x_2}
\over x_1-x_3}+{\Theta\over (x_1-x_2)(x_1-x_3)},
$$
\begin{equation}
\bar X^2=\Theta_{x_1x_3}+{\Theta_{x_3}\over x_1-x_2} +{\Theta_{x_1}
\over x_2-x_3}+{\Theta\over (x_2-x_1)(x_2-x_3)}, \label{A4}
\end{equation}
$$
\bar X^3=\Theta_{x_2x_1}+{\Theta_{x_2}\over x_3-x_1} +{\Theta_{x_1}
\over x_3-x_2}+{\Theta\over (x_3-x_2)(x_3-x_1)},
$$
where $\Theta=\Theta^1(x_2x_3)+\Theta^2(x_1x_3)+\Theta^3(x_1x_2)$ and
$\Theta^i$ are arbitrary functions of two independent arguments. So it
is possible to state that $\Theta$ is arbitrary solution of the equation
$$
\Theta_{x_1x_2x_3}=0
$$
Of course, (\ref{A4}) is equivalent to one represented in the main text and 
used as an initial guess to find the general solution in the form (\ref{27}).

\end{document}